# Controlling The Contrast Between Absorptivity and Emissivity in Nonreciprocal Thermal Emitters

Sina Jafari Ghalekohneh[1], Changkang Du[1], and Bo Zhao[1,a]

[1]Department of Mechanical Engineering,

University of Houston, Houston, 77204, USA

## ABSTRACT

Recent advancements in nonreciprocal thermal emitters challenge the conventional Kirchhoff's law, which states that emissivity and absorptivity should be equal for a given direction, frequency, and polarization. These emitters can break Kirchhoff's law and enable unprecedented thermal photon control capabilities. However, current studies mainly focus on increasing the magnitude of the contrast between emissivity and absorptivity, with little attention paid to how the sign or bandwidth of the contrast may be controlled. In this work, we show such control ability can be achieved by coupling resonances that can provide opposite contrasts between emissivity and absorptivity.

[a] Contact information: Tel: (713) 743-2557, E-mail: bzhao8@uh.edu

As a fundamental symmetry of electromagnetic waves[1-5], reciprocity governs a class of photonic transport properties. For thermal emitters, the role of reciprocity is described by Kirchhoff's law[6-14], which states that the emissivity ($\varepsilon$) and absorptivity ($\alpha$) are equal for a given direction, frequency, and polarization. However, as shown in recent studies, nonreciprocal emitters can violate Kirchhoff's law and allow $\varepsilon$ and $\alpha$ to be separately controlled[11,13,15-17]. Such functionality leads to new ways to control heat radiation[13,15,18-20], record-breaking solar energy harvesting technologies[19,21-24], and mechanical proportion with radiative photons[25,26].

For nonreciprocal thermal emitters, a key figure of merit is the contrast between $\alpha$ and $\varepsilon$, defined as $\eta = \alpha - \varepsilon$. Reciprocal emitters follow Kirchhoff's law and are guaranteed to have $\eta = 0$. For nonreciprocal emitters, $\eta$ is nonzero and can vary between $-1$ to 1. The magnitude of the contrast $|\eta|$ indicates degree of violation of Kirchhoff's law, whereas the sign of $\eta$ directly controls the photon transport behavior, for example, net photon flow direction in heat circulators[23,27]. In heat circulators, one emitter should absorb heat in one direction without emitting it in that direction, and exhibit the opposite behavior in the other direction, which means $\eta$ should have different sign for these two directions. Most of the nonreciprocal thermal emitter designs so far focus on optimizing $|\eta|$. One can now achieve large $|\eta|$ over a fairly broad[28,29] or narrow[30,31] band using various readily-achievable magnetic effect. However, little has known on how the sign of $\eta$ can be controlled.

In this Letter, we propose a method to control the sign of $\eta$ by coupling two modes that can provide opposite contrasts between emissivity and absorptivity, and we implement temporal coupled-mode theory to support the proposed approach. In a resonant system that is open and interacts with propagation waves, the temporal coupled mode theory provides a phenomenological description of the dynamics of the resonant system. The dynamic behavior obtained based on this description align closely with rigorous numerical simulations of resonant systems, and it has found extensive use as a guiding principle in the design of optical devices[32].

According to temporal coupled-mode theory[33], the spectral directional absorptivity and emissivity of an emitter are respectively given by[17]

$$\alpha_m = \frac{2\gamma_i \kappa_m^* \kappa_m}{(\omega-\omega_0)^2-(\gamma_i+\gamma_r)^2}, \quad (1)$$

and

$$\varepsilon_m = \frac{2\gamma_i d_m^* d_m}{(\omega-\omega_0)^2-(\gamma_i+\gamma_r)^2}, \quad (2)$$

where $m$ is the label for the channel in the direction of interest, $\omega_0, \gamma_i$, and $\gamma_r$ are the resonant frequency, intrinsic decay rate due to material loss, and radiative decay rate of the resonance, respectively, and $\kappa_m$ and $d_m$ are the in-coupling and out-coupling rate of channel $m$, respectively. In reciprocal systems, $\kappa_m$ and $d_m$ are identical, which yields $\eta = 0$. In nonreciprocal systems, the difference between $\kappa_m$ and $d_m$ directly results in $\eta$

$$\eta = \frac{2\gamma_i(\kappa_m^*\kappa_m - d_m^* d_m)}{(\omega-\omega_0)^2-(\gamma_i+\gamma_r)^2}. \quad (3)$$

One therefore can control the sign of $\eta$ through designing $d_m^* d_m$ and $\kappa_m^* \kappa_m$. We accordingly propose systems that contain two resonances with one of them possesses $d_m^* d_m > \kappa_m^* \kappa_m$ whereas the other one $d_m^* d_m < \kappa_m^* \kappa_m$. We show that the coupling of the two resonances can be used to control the sign of $\eta$ through manipulating the decay rate of these resonances to the same channel. One can even annihilate the contrast and achieve $\eta = 0$, creating reciprocal behaviors in nonreciprocal emitter systems.

We start with an exemplary system containing magnetic Weyl semimetals. As a kind of topological materials, magnetic Weyl semimetals possesses intrinsic time-reversal-symmetry breaking effect, which gives a nonreciprocal behavior as discussed in several prior work[15,16,19,34-36].

In contrast to magneto-optical materials which require an external magnetic field to achieve nonreciprocity[13], the nonreciprocal effect of magnetic Weyl semimetals is intrinsic and does not require external magnetic field. The momentum separation of the Weyl cones, 2**b**, plays a similar role as an applied magnetic field for magneto-optical systems. As for the dielectric function of magnetic Weyl semimetals, without losing generality, we consider a case[15] when **b** is along the $z$ direction $\mathbf{b} = b\hat{\mathbf{k}}_z$

$$\bar{\bar{\varepsilon}} = \begin{bmatrix} \varepsilon_d & i\varepsilon_a & 0 \\ -i\varepsilon_a & \varepsilon_d & 0 \\ 0 & 0 & \varepsilon_d \end{bmatrix}, \quad (4)$$

where

$$\varepsilon_a = \frac{be^2}{2\pi^2 \hbar \omega}. \quad (5)$$

When $b \neq 0$, $\varepsilon_a \neq 0$ and $\bar{\bar{\varepsilon}}$ becomes asymmetric that breaks reciprocity[33]. The diagonal element $\varepsilon_d = \varepsilon_b + i\frac{\sigma}{\omega\varepsilon_0}$, where $\varepsilon_b$ is the background permittivity and $\sigma$ is the bulk conductivity obtained from

$$\sigma = \frac{\varepsilon_0 r_s g E_F}{6\hbar} \Omega G\left(\frac{E_F \Omega}{2}\right) + i \frac{\varepsilon_0 r_s g E_F}{6\pi\hbar} \left\{ \frac{4}{\Omega}\left[1 + \frac{\pi^2}{3}\left(\frac{k_B T}{E_F(T)}\right)^2\right] + 8\Omega \int_0^{\xi_c} \frac{G(E_F \xi) - G\left(\frac{E_F \Omega}{2}\right)}{\Omega^2 - 4\xi^2} \xi d\xi \right\}. \quad (6)$$

The first part of the second term corresponds to the intraband transition and has Drude-like form[37]. $\Omega = \hbar(\omega + i\tau^{-1})/E_F$ is the normalized complex frequency, $\tau^{-1}$ is the scattering rate, $E_F$ is the chemical potential, $T$ is the temperature and we assume is 300 K for this study, $k_B$ is the Boltzmann constant and $G(E) = n(-E) - n(E)$, where $n(E)$ is the Fermi-Dirac distribution function. $r_s = e^2 / (4\pi\varepsilon_0 \hbar v_F)$ is the effective fine structure constant, $e$ is the elementary charge, $\varepsilon_0$ is the permittivity of vacuum, $\hbar$ is the reduced Planck constant, $v_F$ is the Fermi velocity, $g$ is the number of Weyl points, and $\xi_c = E_C/E_F$ is the normalized cutoff energy, where $E_C$ is the cutoff energy beyond which the band dispersion is no longer linear.

The thermal emitters considered in this work all belong to the black/white symmetry group, where the absorptivity and emissivity can be obtained based on reflectivity and transmissivity as[38]

$$\alpha(\theta) = 1 - R(\theta) - T(\theta), \tag{7}$$

and

$$\varepsilon(\theta) = 1 - R(-\theta) - T(-\theta), \tag{8}$$

where $\theta$ is the angle of incidence as shown in the inset of Fig. 1a, $R$ is the reflectivity and $T$ is the transmissivity. That is, the absorptivity in direction $\theta$ is computed from the $R$ and $T$ of the same direction, whereas the emissivity in direction $\theta$ is computed from the $R$ and $T$ in direction $-\theta$.

As shown in Fig. 1a, the emitter consists of a single layer Weyl semimetal with a thickness of 50 nm, on a silver substrate. We consider transverse magnetic (TM) waves with an electric field in the *x–y* plane. The system is in a Voigt geometry with the separation of the Weyl nodes along the *z*-direction and plane of incidence being the *x-y* plane. We focus on the mid-infrared range, in which the properties of magnetic Weyl semimetals are dominated by the free electron Drude contribution. For a chemical potential $E_F = 0.2$ eV, the real part of $\varepsilon_d$ crosses zero near 6 μm[39], and a Weyl semimetal film can support leaky modes called Brewster modes[20,40], which can greatly enhance the nonreciprocal radiative properties, as reported in several prior works[28,41]. Figure 1a illustrates a contour of $\eta$ for the single layer Weyl semimetal emitter on a silver substrate. The nonreciprocal behavior is greatly enhanced near 6 μm where Brewster modes are excited. The sign of $\eta$ is positive (red color) for most of wavelengths and angular range considered here, indicating that $d_m^* d_m < \kappa_m^* \kappa_m$.

To justify this, we obtain $\kappa_m$ and $d_m$ from the reflectivity given by coupled mode theory, $R_{nm} = \left| C_{nm} + d_n \frac{\kappa_m}{i(\omega-\omega_0)+\gamma_i+\gamma_r} \right|^2$, where $m$ is the index of the channel for the incoming light, $n$ is the index of the channel for the reflected light, and $C$ is the background scattering coefficient. Implementing the above equation, we obtain $d$ and $\kappa$ using the calculated reflection spectrum at 85 degree as listed in Table 1. The coupled-mode theory predictions agree with the computed reflection spectrum, as shown in Fig. 2a. Indeed, we have $d_m^* d_m < \kappa_m^* \kappa_m$ for this Brewster mode. This means that in the studied wavelength range, the out-coupling rate contributing to emissivity is smaller than the in-coupling rate contributing to absorptivity. This observation is supported by Fig. 1a, where η is positive in the dominant wavelength range for an angle of incidence of 85 degrees. This is also true for other Brewster modes across the wavelength and angular range shown in the contour. Such properties are not readily tunable by changing the thickness of the film, as shown in Fig. 1b, where we double the thickness of the film.

Our idea is to use mode coupling such that the sign of $\eta$ can be controlled as needed for different wavelengths. To do that, we start by introducing another mode with $d_m^* d_m > \kappa_m^* \kappa_m$. As a readily available option, we can use the same system but flip the direction of 2**b** to be along -$z$ direction as shown in Fig. 1c. In doing so, $R(\theta)$ and $R(-\theta)$ flips as compared to the system shown in Fig. 1a, and therefore, as can be seen in Table 1 and Fig. 2b, $d$ and $\kappa$ of this Brewster mode switches, yielding $d_m^* d_m > \kappa_m^* \kappa_m$. In such system, the overall shape of the contour remains the same as the system of Fig. 1a with the sign of $\eta$ flipped, but $\eta$ still lacks tunability.

Table 1- The values of the coupled-mode theory parameters (in-coupling and out-coupling rates) for the incident angle of 85 degree

| Direction of 2**b** | $\kappa\kappa^*$ | $dd^*$ |
|---|---|---|
| $z$ direction | 0.057 | 0.04 |
| $-z$ direction | 0.04 | 0.057 |

We propose an emitter shown in Fig. 1d that hosts the abovementioned two Brewster modes with opposite contrasts. As shown by the contour of $\eta$ in Fig. 1d, the coupling of the two modes can effectively alter the sign of $\eta$ near 6.3 μm and make $\eta$ positive in 6.2 μm $< \lambda <$ 6.4 μm and angular range larger than 65 degree. Therefore, the coupling enables the control of sign of $\eta$ independently in this wavelength and angular ranges. Besides controlling the sign of $\eta$, the coupling of the two modes also enhances the magnitude of $\eta$ in smaller angles around 6 μm, as shown in Fig. 1d.

In between the wavelength ranges with opposite $\eta$, there is a region where $\eta$ approaches to zero, as indicated by the white band between the red and blue region. Within this white band, the emissivity and absorptivity are approximately equal, retrieving a reciprocal behavior. This band therefore serves as a boundary for nonreciprocal properties in the angular and wavelength domain in nonreciprocal systems. One can control the location of the white region around which a sign flip of $\eta$ can be expected, and therefore control the bandwidth of nonreciprocal regions within which $\eta$ possesses the same sign.

To further illustrate the bandwidth tuning capability through mode coupling, we introduce a system[28] that can support broadband nonreciprocal properties. As shown in Fig. 3a, the emitter consists of a silver (Ag) substrate covered by ten layers InAs, a magneto-optical material, with slightly different doping levels for each layer. We consider TM waves and an applied magnetic field of 1.5 T along the -z-axis. In doing so, similar to Weyl semimetals, each layer of InAs can support a Brewster mode. Since the doping levels are different, the Brewster modes supported by each layer are also slightly different in wavelength. With doping levels gradually increasing from top to bottom[29], the multilayer system can produce nonreciprocal behavior in a broad wavelength range, as theoretically and experimentally[28] demonstrated recently. Here, we consider a case where the thickness of each layer is 350 nm and the doping level of each layer increases linearly from $4 \times 10^{17} \text{cm}^{-3}$ (for the first layer) to $8.5 \times 10^{17} \text{cm}^{-3}$ (for the last layer). As depicted in Fig. 3a, $\eta$ is indeed not zero and remains negative in a large continuum region in the wavelength and angle space. The design provides an effective platform for nonreciprocal properties, but $\eta$ lacks tunability in the continuum.

We introduce a dielectric layer beneath the multi-layered InAs, as illustrated in Fig. 3b, to enable the mode coupling necessary to control $\eta$. We assume the dielectric layer to have a dielectric constant of 16, which is close to the property of Germanium in this wavelength range. When the thickness of the dielectric layer is much smaller than the wavelength, the magnitude of $\eta$ can be enhanced as reported in Ref.[42]. Here, we focus on enabling mode coupling to control $\eta$ with a much thicker dielectric, rather than enhancing the magnitude of $\eta$. An immediate consequence of this dielectric layer is the creation of bright bands in the $\eta$ contour. As discussed in the Weyl semimetal case, the bright bands imply that the new introduced modes offset the effects of the broadband Brewster modes on the decay rates. These new modes are the Fabry-Perot modes associated with the dielectric layer. They couple with the Brewster mode and alter the spectral radiative properties of the emitter. Depending on their magnitude, they can either nullify η or reverse its sign. The dispersion of the Fabry-Perot modes is[43] $1 - r_1 r_2 e^{i2\beta} = 0$, where $r_1$ and $r_2$ are the reflection coefficients from dielectric layer to air and silver, respectively, and $\beta = k_y d$ represents the phase shift encountered while propagating within the dielectric layer with thickness $d$. $k_y$ is the wavevector of the propagating wave along $y$-axis in the dielectric layer. In solving the dispersion, the InAs layer stack is modeled as a single layer with a thickness of 3500 nm and a constant doping level equal to the average of the doping levels of all layers. The dispersion of the Fabry-Perot modes is overlayed in Fig. 3b, which agrees well with the white bands in the $\eta$ contour, justifying the excitation of Fabry-Perot modes.

The excitations of Fabry-Perot modes bring new tunability to the broadband nonreciprocal properties. $\eta$ for different wavelengths can be now separately controlled, as well as the wavelength range that has the same sign of $\eta$. These tunability can be also flexibly adjusted through controlling the Fabry-Perot modes. For instance, by increasing the thickness of the dielectric layer of the structure shown in Fig. 3b, more orders of Fabry-Perot modes can be introduced to the nonreciprocal continuum, as illustrated in Fig. 3c. In doing so, less area in the contour will have a blue color, and therefore the Fabry-Perot modes effectively offset the nonreciprocal effect of the broadband Brewster modes when integrated over the whole wavelength range. Such functionality can be critical to applications such as heat circulators and persistent heat current systems[27]. One could also achieve similar effect through decreasing the quality factor of the Fabry-Perot modes by, for example, increase the material loss in the dielectric layer. In Fig. 4, we show the results when the dielectric function is artificially changed to 16+1i, where adding loss decreases the quality factor of Fabry-Perot resonance. In doing so, Fabry-Perot resonances offset the nonreciprocal effect of Brewster modes in a wide angular range compared to Fig. 3b, as can be observed in Fig. 4.

In conclusion, we demonstrate the capability offered by mode coupling in controlling the nonreciprocal radiative properties. By coupling modes that can yield opposite signs for $\eta$, one can effectively control the bandwidth, the direction, the sign, and also the magnitude of the contrast between absorptivity and emissivity. Extensive efforts are ongoing to increase the magnitude of $\eta$ in nonreciprocal emitters. The integration of these efforts with the methodology described in this study can significantly enhance the ability to fine tune nonreciprocal radiative properties for a wide range of applications, representing a substantial step forward in nonreciprocal thermal photonics.

## ACKNOWLEDGEMENTS

The authors acknowledge the start-up funding from the University of Houston and the funding from National Science Foundation under grant no. CBET-2314210.

## AUTHOR DECLARATIONS

### Conflict of Interest

The authors have no conflict of interest to disclose.

### Author Contributions

**Sina Jafari Ghalekohneh**: Formal analysis (lead); Investigation (lead); Methodology (lead); Software (lead); Validation (lead); Visualization (lead); Writing – original draft (lead). **Changkang Du**: Formal analysis (supporting); Investigation (supporting); Methodology (supporting); **Bo Zhao**: Conceptualization (lead); Formal analysis (lead); Investigation (lead); Methodology (lead); Project administration (lead); Resources (lead); Supervision (lead); Writing – original draft (lead); Writing – review & editing (lead).

## DATA AVAILABILITY

The data that support the finding of this study are available from the corresponding author upon reasonable request.

**Figure Captions**

Fig. 1. The contrast between absorptivity and emissivity of (a) thin-film Weyl semimetal layer on a silver substrate with the Weyl nodes separation in the *z* direction (pink), (b) the similar structure to (a) but the thickness of Weyl semimetal layer is doubled, (c) thin film Weyl semimetal layer on a silver substrate when the Weyl nodes separation is in the *-z* direction (orange), and (d) double layer of Weyl semimetal when the top layer is the Weyl semimetal shown in (a) and bottom layer is the structure shown in (c).

Fig. 2 (a) The reflectivity of the structure shown in Fig. 1a and (b) the reflectivity of the structure shown in Fig. 1c at angle of incidence of 85 degree. For both cases, the coupled-mode theory predictions are overlaid on the graphs using star symbols.

Fig. 3 (a) The contrast between absorptivity and emissivity of a multi-layer InAs with different doping levels. The lighter shades represent lower levels of doping and darker represents higher. (b) The contrast between absorptivity and emissivity when a 10-μm Ge layer is inserted, and (c) a 20-μm Ge layer is inserted. The solution of Fabry-Perot dispersion is shown with dashed black line.

Fig. 4 The contrast between absorptivity and emissivity of a multi-layer InAs with different doping levels with a dielectric layer of 10 μm and permittivity 16+1i.

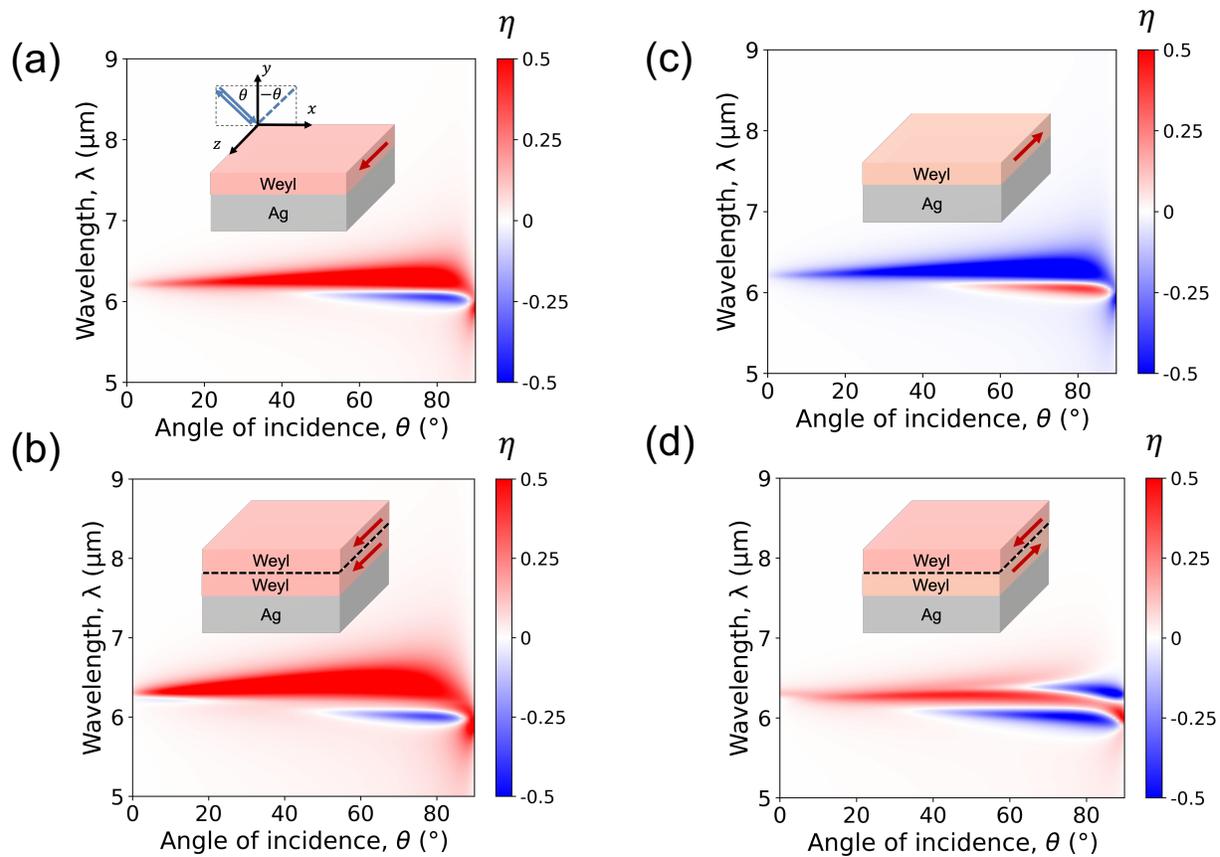

Figure 1

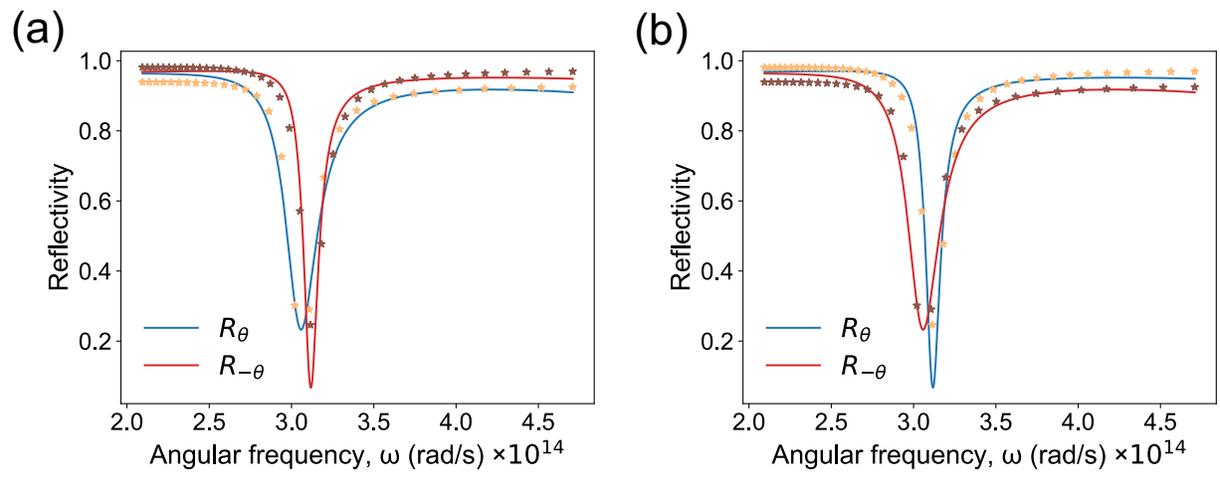

Figure 2

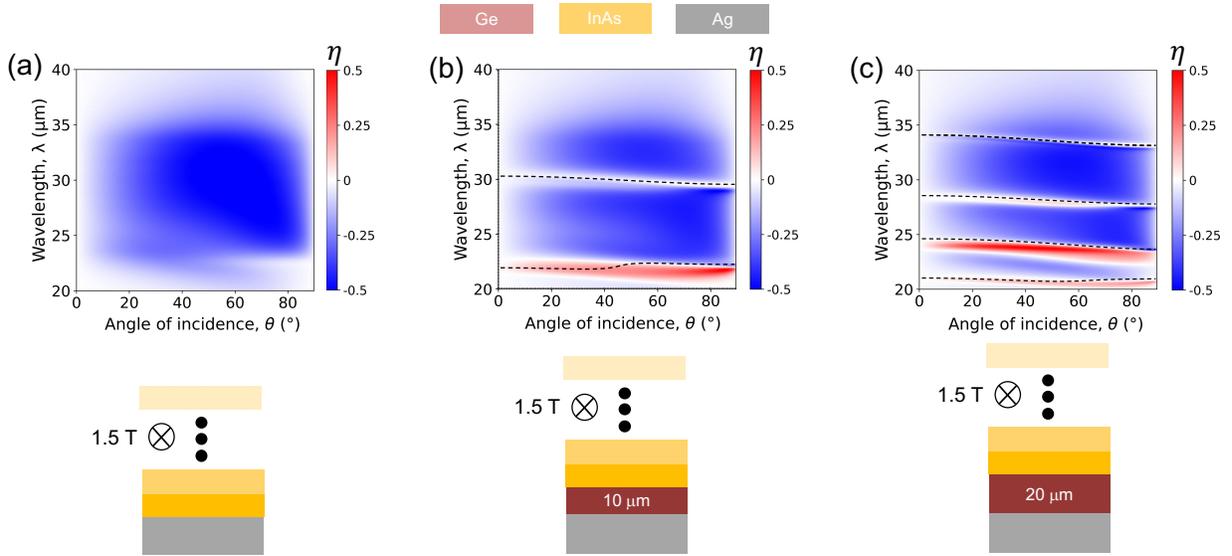

Figure 3

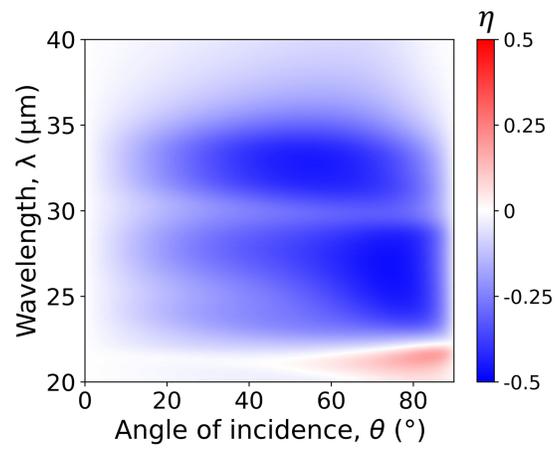

Figure 4